# Effect of structural distortion and nature of bonding on the electronic properties of defect and Li-doped $CuInSe_2$ Chalcopyrite Semiconductors

S.Mishra, B.Ganguli[*]

*National Institute of Technology, Rourkela-769008, Odisha, India*

**Abstract**

We report the structural and electronic properties of chalcopyrite semiconductors $CuInSe_2$, $CuIn_2Se_4$ and $Cu_{0.5}Li_{0.5}InSe_2$. Our calculation is based on Density Functional Theory within tight binding linear muffin-tin orbital (TB-LMTO) method. The calculated lattice constants, anion displacement (u), tetragonal distortion ($\eta$ = c/2a) and bond lengths agree well with experimental values. Our result shows these compounds are direct band gap semiconductors. Our calculated band gaps, 0.79eV and 1.08 eV of $CuInSe_2$ and $Cu_{0.5}Li_{0.5}InSe_2$ respectively agree well with the experimental values within the limitation of LDA. The band gap of $CuIn_2Se_4$ is found to be 1.50 eV. The band gap reduces by 59.57%, 23.61% and 48.82% due to p-d hybridization and reduces by 16.85%, 9.10% and 0.92% due to structural distorion for $CuInSe_2$, $CuIn_2Se_4$ and $Cu_{0.5}Li_{0.5}InSe_2$ respectively. We also discuss the effect of bond nature on electronic properties of all three compounds.

[*]corresponding author. Tel.: +91661 2462725; fax: +91661 2462999
   *Email address:* biplabg@nitrkl.ac.in (B.Ganguli)

## 1. Introduction

Among the $A^I - B^{III} - C_2^{VI}$ compounds, $CuInSe_2$ having band gap energy 1.04 eV [1] is regarded as one of the best investigated material for thin film solar cells. But partial Li-doping in place of Cu atoms not only changes its physyical properties but also tailor the energy band gap suitable for optimum conversion efficiency for solar cells. The defect chalcopyrites have vacancies at the cation sites in such a manner that they do not break translational symmetry. Due to the defect structure the compounds are porous. Because of porosity these systems have attracted special attention of the physics community. Various type of impurities including magnetic impurities can be doped into the vacancies to design new class of materials, like Dilute magnetic semiconductors (DMS) for spinstronics application [2], optoelectronic devices[3] and solar cells [4]. The presence of vacancy, and more than two type of atoms provides desired band gap, electronic and optical properties to attain a maximum criteria for new emerging functional materials.

In this communication we investigate a comparative study of the structural, electronic properties and bond nature of $CuInSe_2$, $CuIn_2Se_4$ and $Cu_{0.5}Li_{0.5}InSe_2$ compounds. There are very few experimental studies carried out for $CuIn_2Se_4$ [5] and $Cu_{0.5}Li_{0.5}InSe_2$. and no detailed theoretical study has been reported till date. Mitaray et. al. [6] studied the growth, structure and optical properties of $Cu_{1-x}Li_xInSe_2$ thin films. They found that $Cu_{0.5}Li_{0.5}InSe_2$ has chalcopyrite phase and it is a direct band gap semiconductor for $x \leqslant 0.6$. Weise et.al. [7] discussed the preparation and the structural phase transformation of $Cu_{1-x}Li_xInSe_2$. They also reported this to be chalcopyrite structure below $810^0C$ for $0 \leqslant x \leqslant 0.55$. Extensive experimetal [8, 9, 10, 11] and theoretical studies [12, 13] have been carried out for $CuInSe_2$ compounds. We have chosen this already much studied system to validate our methodology and calculation and extend the study to other two systems which have not been much studied. Our main motivation is to study the effect of structure and p-d hybridization on the electronic properties. We have also discused the bond nature of these compounds. In our earlier work on pure chalcopyrite semiconductors [14] and defect chalcopyrite systems [15] we have shown that due to the presence of group I element (Cu,Ag), d-orbital contribution is very prominent. The main contribution of d-orbitals to upper valence band is significant which affects band gap. In case of $Cu_{0.5}Li_{0.5}InSe_2$, though Lithium has no d-orbital contribution, its doping has significant effect on Cu d and Se p hybridization. For structural properties, we calculate the lattice parameters, tetragonal distortion, anion displacement parameters and bond lengths by energy minimization proceedure. We also calculate the bulk modulus for $CuIn_2Se_4$ using extended Cohens formula [16] and we have extended this formula to calculate bulk modulus for $Cu_{0.5}Li_{0.5}InSe_2$. There have been no calculation so far of bulk modulus of these two compounds. For our study we have used highly successful Density Functional Theory (DFT) based first principle technique, tight binding linearized muffin-tin Orbital (TB-LMTO) method. In TB-LMTO method, the basis functions are localized. Therefore, very few basis functions are required to represent the highly localized d-orbital of Cu in the systems under study. Hence the calculation is not only cost effective, it gives also the accurate result.

## 2. Methodology

The ab-initio method is based on Density Functional Theory of Kohn-Sham [17]. The one electron energy is given by Khon-Sham equation.

$$\left[-\nabla^2 + V_{eff}(\mathbf{r})\right]\psi_i(\mathbf{r}) = \varepsilon_\mathbf{i}\psi_\mathbf{i}(\mathbf{r}) \tag{1}$$

where the effective potential,

$$V_{eff}(\mathbf{r}) = 2\int dr' \frac{\rho(\mathbf{r}')}{|\mathbf{r}-\mathbf{r}'|} - 2\sum_R \frac{Z_R}{|r-R|} + \frac{\delta E_{XC}[\rho]}{\delta\rho(\mathbf{r})} \tag{2}$$

The total electronic energy is a function of electron density which is calculated using variational principle. This requires self consistent calculations. In practice the Kohn-Sham orbitals $\psi_i(r)$ are usually expanded in terms of some chosen basis function. We use the well established TB-LMTO method, discussed in detail elsewhere [18, 19] for the choice of the basis function. Electron correlations are taken within LDA of DFT [17, 20]. We have used the von Barth-Hedin exchange [21] with 512 $\mathbf{k}$-points in the irreducible part of the Brillouin zone. The basis of the TB-LMTO starts from the minimal set of muffin-tin orbitals of a KKR formalism and then linearizes it by expanding around a 'nodal' energy point $E_{\nu\ell}^\alpha$. The wave-function is then expanded in this basis :

$$\Phi_{j\mathbf{k}}(\mathbf{r}) = \sum_\mathbf{L}\sum_\alpha \mathbf{c}_{\mathbf{L}\alpha}^{\mathbf{jk}} \left[\phi_{\nu\mathbf{L}}^\alpha(\mathbf{r}) + \sum_{\mathbf{L}'}\sum_{\alpha'} \mathbf{h}_{\mathbf{LL}'}^{\alpha\alpha'}(\mathbf{k})\dot{\phi}_{\nu\mathbf{L}'}^{\alpha'}(\mathbf{r})\right] \tag{3}$$

where,

$$\begin{aligned}
\phi_{\nu L}^{\alpha}(\mathbf{r}) &= \imath^{\ell}\, Y_L(\hat{r})\, \phi_{\ell}^{\alpha}(r, E_{\nu\ell}^{\alpha}) \\
\dot{\phi}_{\nu L}^{\alpha}(\mathbf{r}) &= \imath^{\ell}\, Y_L(\hat{r})\, \frac{\partial \phi_{\ell}^{\alpha}(r, E_{\nu\ell}^{\alpha})}{\partial E} \\
h_{LL'}^{\alpha\alpha'}(\mathbf{k}) &= (C_L^{\alpha} - E_{\nu\ell}^{\alpha})\, \delta_{LL'}\delta_{\alpha\alpha'} + \sqrt{\Delta_L^{\alpha}}\, S_{LL'}^{\alpha\alpha'}(\mathbf{k})\, \sqrt{\Delta_{L'}^{\alpha'}}
\end{aligned}$$

$C_L^{\alpha}$ and $\Delta_L^{\alpha}$ are TB-LMTO potential parameters and $S_{LL'}^{\alpha\alpha'}(\mathbf{k})$ is the structure matrix.

## 3. Result and discussion

### 3.1. Structural properties :

The tetragonal unit cell of a typical chalcopyrite semiconductor consists of two Zinc blende unit cells. There are two types of cations in the unit cell. In a defct chalcopyrite there are 50% vacancy in one type of cation compared to pure chalcopyrite. In Li-doped $CuInSe_2$ there are three types of cations in a unit cell unlike pure and defect chalcopyrites. One unit cell of $Cu_{0.5}Li_{0.5}InSe_2$ is shown in figure 1. In $Cu_{0.5}Li_{0.5}InSe_2$, the 50% vacancy in Cu-atom position is occupied by Li-atoms. In both defect and Li-doped $CuInSe_2$, the vacancies/Li atoms are occupied in such a manner that they do not break translational symmetry. There are two Cu atoms, two Li atoms, four In atoms and eight Se atoms per unit cell. The positions of the various atoms in the tetragonal unit cell of $CuInSe_2$ are : Cu : 0.0 0.0 0.0; In: 0.0 0.0 0.5; Se: u 0.25 0.125, where 'u' is anion displacement parameter. For the defect and Li-doped $CuInSe_2$ the positions of the various atoms in the tetragonal unit cell are : Cu : 0.0 0.0 0.0; Vacancy/Li : 0.0 0.5 0.75; In1 : 0.0 0.0 0.5; In2 : 0.0 0.5 0.25; Se: $u_x\ u_y\ u_z$ where '$u_x$', '$u_y$' and '$u_z$' are anion displacement parameters along three axes. Unit cell with $\eta = 1$, $u_x$, $u_y$ and $u_z$ equal to 0.25, 0.25 and 0.125 respectively is refered as ideal case. This is defined in analogous to binary zinc-blende (ZnS) structure [12]. In $CuIn_2Se_4$ system, each Se atom has one Cu cation, two In cation and one vacancy as nearest neighbors. Due to different atoms and one vacancy as neighbors, Se atom acquires an equlibrium position closer to the vacancy than to the other three cations. Where as in case of $Cu_{0.5}Li_{0.5}InSe_2$, as shown in figure 1, each Se atom has one Cu cation, two In cations and one Li atom as nearest neighbors. The Se atom moves towards the Li atom to acquire the equilibrium position. This new position of anion is called anion displacement. In both defect and doped $CuInSe_2$, Se atoms shift along all the three directions unlike only along x-direction as found in the case of $CuInSe_2$ chalcopyrites [12]. This is due to the reduction in symmetry in case of $CuIn_2Se_4$ and $Cu_{0.5}Li_{0.5}InSe_2$ system. Therefore all cations-Se bond lengths are inequivalent in defect and doped $CuInSe_2$ where as this is not the case for $CuInSe_2$ [12] (table 1).

For self consistent calculation, we introduce empty spheres because the packing fraction is low due to tetrahedral co-ordination of ions. We ensure proper overlap of muffin tin spheres for self consistency and the percentage of overlap is found. Table 1 shows the calculated structural parameters 'a', 'c', tetragonal distortion, anion displacement and bulk modulus (B). These parameters are found by energy minimization proceedure. We calculate the bulk modulus 'B' for $CuIn_2Se_4$ using extended Cohen formula [16]. The bulk modulus of $Cu_{0.5}Li_{0.5}InSe_2$ is calculated by the slightly modified form of extended

Cohen formula given in equation (4).

$$B = \frac{1971 - 220\lambda}{4} \sum_{i=1,2,3,4} \frac{1}{d_i^{3.5}} \quad (4)$$

where B is in GPa and the nearest-neighbor distances $d_i$ in $A^0$. The distances $d_i$ in equation (4) are the bond lengths of Cu-Se, Li-Se, In1-Se, In2-Se. The ionicity coefficient $\lambda$ is taken equal to 2, analogous to II-VI semiconductors. Structural parameters and bulk modulus are listed in table 1. Calculated bond lengths are listed in table 5.

*3.2. Electronic properties*

(i) $CuInSe_2$ : Our result shows this compound a direct band gap semiconductor. Total density of states (TDOS) (figure 2a) shows four major sub valence bands of different band widths. The first two subbands below the valence band maximum have band widths 1.9 eV and 2.7 eV respectively. They are separated by very narrow band gap of 0.95 eV . The third and fourth subbands have band width 0.8 eV and 1.3 eV respectively. They are separated by a large band gap of 6.4 eV. The second and third subbands are separated by a very narrow band gap of 0.3 eV. The major contribution for the formation of the first subband comes from Cu d orbitals and weak contribution from Se p orbitals. Where as it is other way round for the formation of the second subband. Contribution for the formation of the third sub band comes from the admixture of In s and Se p orbitals. The lowest subband is Se 4s band. Figure 2(b) shows Partial density of states (PDOS) for Cu-d and Se-p. It is clear from the figure that Cu d and Se p hybridization contribute to upper valence band near fermi level and there is no contribution of Cu d states to conduction band. The main contribution to the conduction band comes from Cu p, In p and Se p states and very weak contribution from Cu s and In s orbitals. The conduction band width is 16.0 eV.

(ii) $CuIn_2Se_4$ : Unlike $CuInSe_2$, this compound is slightly p-type semiconductor. TDOS (figure 3(a)), shows three sub valence bands of different band widths. The first two subbands have band widths 4.6 eV and 1.0 eV respectively and are separated by very narrow band gap of 0.4 eV. The lowest subband with band width 1.3 eV is Se 4s band. There is a large band gap of nearly 6.1 eV between the lowest and second subband. Contribution for the second subband comes from the admixture of In s and Se p states. Calculated PDOS (figure 3(b)) shows that the upper valence band is dominated by Cu d and Se p hybride orbitals. This semiconductor shows p-type conductivity because Se p orbitals crosses fermi level. The conduction band width is approximately 14.9 eV. The main contribution comes from Cu p, In p and Se p states and a very weak contribution from Cu s and In s states for the formation of the conduction band.

(iii) $Cu_{0.5}Li_{0.5}InSe_2$ : The band structure and TDOS (figure 4) show four major sub valence bands of different band widths. The two upper most subvalence bands have band widths of 1.2 eV and 3.1 eV respectively. They are separated by 0.4 eV. Calculated PDOS (figure 5(b)) of Cu d and Se p orbitals show that the main contribution to the upper most sub valence band comes from Cu d orbital and very weak contribution comes from Se p orbitals. The second sub valence band is mainly formed due to the contribution of Se p and very weak contribution from Cu d states like in the case of $CuInSe_2$. The lowest band of band width 1.2 eV is formed due to the contribution of Se 4s states. The third sub valence band is formed due to the admixture of In s and Se p orbitals. The main contribution to conduction band comes from Cu p, Se p, In p and very weak contribution comes from In s and Li s orbitals. From PDOS of Li s and Li p orbitals (figure 5(a)), we do not find any significant participation of Li-orbitals neither in valence band nor in conduction band. The conduction band width is found to be 15.2 eV.

In all the cases, valence band maximum (VBM) and conduction band minimum (CBM) are located at the center of Brillouin zone denoted as 'G'($\Gamma$ point). This indicates that they are all direct band gap compounds. Experimental and our calculated band gaps are listed in table 2. It is known that LDA underestimates band gap by 30-50% [23]. Within this LDA limitation, our results are in good agreement with experimental band gap. No detailed theoretical study of band gap for $CuIn_2Se_4$ and $Cu_{0.5}Li_{0.5}InSe_2$ are reported in literature.

*3.3. Effect of p-d hybridization on electronic properties*

It is known that p-d hybridization has significant effect on the band gap in the case of Cu based compounds like $CuInSe_2$[12]. To see this effect explicitly, we calculate the TDOS without the contribution of the d-orbitals for ideal $CuInSe_2$, $CuIn_2Se_4$ and $Cu_{0.5}Li_{0.5}InSe_2$ systems. Therefore we first freeze d-electrons and treat these electrons as core electrons. Figures 6 show the TDOS with d-electron of A atoms as frozen for all three systems respectively. We summarize the band gaps with and without contribution of d-electrons of Cu in table 3. The calculated result shows that there is a significant reduction of band gaps in all the cases. The reduction is 59.57% for $CuInSe_2$, 23.61% for $CuIn_2Se_4$ and 48.82% for $Cu_{0.5}Li_{0.5}InSe_2$. The p-d hybridization in chalcopyrite semiconductors can be interpreted on the basis of simple molecular orbital considerations [12]. The p-orbitals that possess the $\Gamma_{15}$ symmetry hybridize with those of the d-orbitals that present the same sysmetry. This hybridization forms a lower bonding state and an upper antibonding state. The antibonding state that constitutes the top of the valence band is predominantly formed by higher energy anion p-states and the bonding state is constituted by the lower energy cation d-states. Perturbation theory [24] suggests that the two states $\Gamma_{15}(p)$ and $\Gamma_{15}(d)$ will repel each other by an amount inversely proportional to the energy difference between p and d states. So this raising of the upper most state causes a gap reduction. But in defect $CuIn_2Se_4$ and Li doped chalcpyrites, there is a reduction in the atomic percentage of Cu relative to that $CuInSe_2$. This reduces the symmetry. So the repulsion between $\Gamma_{15}(p)$ and $\Gamma_{15}(d)$ decreases and the antibonding state is depressed downwards leading to an increase in band gap. Therefore all the Cu defficient defect chalcopyrites have band gaps greater than that the corresponding pure chalcopyrites [14]. Since Li d-orbital does not contribute to the valence band, therefore we expect reduction in band gap due to p-d hybridization in $Cu_{0.5}Li_{0.5}InSe_2$ and $CuIn_2Se_4$ be equal as both have equal concentration of Cu. Contrary to this, our calculation shows band gap reduction in the Li doped $CuInSe_2$, is much more than the defect $CuIn_2Se_4$. On comparing TDOS for $CuIn_2Se_4$ (figure 3(a)) and $Cu_{0.5}Li_{0.5}InSe_2$ (figure 4), we find the conduction band minimum in the case of Li doped compound shifts significantly towards the fermi level compared to the defect system. This shift is due to the shift of Se-p orbitals towards the fermi level (figure (5(b))) in $Cu_{0.5}Li_{0.5}InSe_2$. The valenece band maximum which lies slightly above the fermi level in $CuIn_2Se_4$, shifts at the fermi level in Li doped compound. There is very negligible contribution of Li-orbitals to valence and conduction bands. Therefore we can say Li acts as a catalyst.

### 3.4. Structural effect on electronic properties

Table 4 shows the structural distortion like bond alternation and tetragonal distortion have some effect on the band gap. A close comparision of TDOS for ideal (figure 7) and non-ideal case (figure 2(a), figure 3(a) and figure 4) of all three compounds show distinct differences in the structure in DOS. For example a sharp peak is found at an energy -2.0 eV for nonideal $Cu_{0.5}Li_{0.5}InSe_2$ (figure 4) compared to the corresponding ideal case (figure 7c). The sharp peak comes due to the contribution of Se-p orbitals. But in case of ideal $Cu_{0.5}Li_{0.5}InSe_2$, DOS is high in Cu d orbitals. There are effects on conduction band also. This shows that structural distortion not only decreases the band gap but it has significant effect on overall electronic properties as well. Similar results are also found for $CuInSe_2$ and $CuIn_2Se_4$ systems. For these two compounds we can find significant effect of structural distortion on band gap. The effect of distortion on valence and conduction bands show that structural distortion is also responsible for significant change in optical properties of such semiconductors.

### 3.5. Bond Nature

Our Calculated bond lengths are listed in table 5. In case of $CuInSe_2$ bond lengths of Cu-Se and In-Se agree well with experimental bond lengths mentioned in [12]. In $CuInSe_2$ the covalent bonding character of the Cu-Se bonding is dominant. When covalent character dominates, Se p-Cu d hybridization plays a major role. This causes narrowing down the band gap more in case of $CuInSe_2$ than the other two. This is because Cu-Se bonding is 50% less in defect and Li doped $CuInSe_2$. The Li-Se bonding possess an ionic character because of large electronegativity difference between Li and Se atoms. This ionic character of bonding increases the band gap in $Cu_{0.5}Li_{0.5}InSe_2$.

## 4. Conclusion

Calculations and study of $CuInSe_2$, $CuIn_2Se_4$ and $Cu_{0.5}Li_{0.5}InSe_2$ suggest that these compounds are direct band gap semiconductors with band gaps of 0.79eV, 1.50 eV and 1.08 eV respectively. Our study further shows that electronic properties of these semiconductors significantly depend on the type of hybridization and structural distortion. The calculation is carried out using DFT based TB-LMTO method. We have used LDA for our exchange co-relation functional. Taking into account of the underestimation of band gap by LDA, our result of band gap and structutal properties agree with experimental values. Detail study of TDOS and PDOS shows that p-d hybridization between atom Cu-d and Se-p reduces the band gap. The reduction is 59.57%, 23.61% and 48,82% respectively for $CuInSe_2$, $CuIn_2Se_4$ and $Cu_{0.5}Li_{0.5}InSe_2$ respectively. Li doping in $CuInSe_2$, shifts Se p orbitals in conduction band towards fermi level which eventually reduces the band gap. Decrement of the band gap due to structural distortion is 16.85%, 9.10% and 0.92% in case of $CuInSe_2$, $CuIn_2Se_4$ and $Cu_{0.5}Li_{0.5}InSe_2$ respectively.

## 5. Acknowledgement

This work was supported by Department of Science and Technology, India, under the grant no.SR/S2/CMP-26/2007. We would like to thank Prof. O.K. Andersen, Max Planck Institute, Stuttgart, Germany, for kind permission to use the TB-LMTO code developed by his group.

**Figure captions**

**Figure 1**: One unit cell of doped chalcopyrite semiconductor $Cu_{0.5}Li_{0.5}InSe_2$.

**Figure 2**: (a) TDOS of $CuInSe_2$ (b) PDOS of Cu d and Se p orbitals of $CuInSe_2$

**Figure 3**: (a) TDOS of $CuIn_2Se_4$ (b) PDOS of Cu d and Se p orbitals of $CuIn_2Se_4$.

**Figure 4**: Band structure and TDOS for non-ideal $Cu_{0.5}Li_{0.5}InSe_2$.

**Figure 5**: (a) PDOS of Li s and Li p (b) PDOS of Cu-d and Se-p orbitals in $Cu_{0.5}Li_{0.5}InSe_2$.

**Figure 6**: TDOS of ideal (a) $CuInSe_2$ (b) $CuIn_2Se_4$ (c) $Cu_{0.5}Li_{0.5}InSe_2$ without hybridization.

**Figure 7**: TDOS for ideal and with hybridization of (a) $CuInSe_2$ (b) $CuIn_2Se_4$ (c) $Cu_{0.5}Li_{0.5}InSe_2$.

Table 1: Calculated structural parameters

| Compounds | $a$ ($\mathring{A}$) | $c/a$ | $a_{exp}$ ($\mathring{A}$) | $c/a_{exp}$ | $u_x(exp)$ | $u_x$ | $u_y$ | $u_z$ | B (GPa) |
|---|---|---|---|---|---|---|---|---|---|
| $CuInSe_2$ | 5.75 | 1.010 | $5.78^a$ | 1.005 | $0.235^a$ | 0.236 | 0.250 | 0.125 | |
| $CuIn_2Se_4$ | 5.74 | 1.014 | | | | 0.251 | 0.235 | 0.126 | 45.44 |
| $Cu_{0.5}Li_{0.5}InSe_2$ | 5.84 | 0.996 | $5.85^b$ | $0.994^b$ | | 0.253 | 0.233 | 0.127 | 60.14 |

$^a$ Ref.[22] $^b$ Ref.[7]

Table 2: % Energy gaps of chalcopyrites.

| Compounds | Experiment | Our work |
|---|---|---|
| | eV | eV |
| $CuInSe_2$ | 1.04 [a] | 0.79 |
| $CuIn_2Se_4$ | 1.04-1.27[b] | 1.50 |
| $Cu_{0.5}Li_{0.5}InSe_2$ | 1.5[c] | 1.08 |

[a] Ref.[1] [b] Ref.[5] [b] Ref.[7]

Table 3: % of Reduction in band gap (eV) due to hybridization for ideal case.

| Systems | With hybridization | Without hybridization | Reduction(%) |
|---|---|---|---|
| $CuInSe_2$ | 0.95 | 2.35 | 59.57 |
| $CuIn_2Se_4$ | 1.65 | 2.16 | 23.61 |
| $Cu_{0.5}Li_{0.5}Se_2$ | 1.09 | 2.13 | 48.82 |

Table 4: Effect of structural distortion on band gap(eV).

| Systems | Ideal | Non-ideal | % of decreament in band gap |
|---|---|---|---|
| $CuInSe_2$ | 0.95 | 0.79 | 16.85 |
| $CuIn_2Se_4$ | 1.65 | 1.5 | 9.10 |
| $Cu_{0.5}Li_{0.5}InSe_2$ | 1.09 | 1.08 | 0.92 |

Table 5: Calculated bond lengths in Å.

| Systems | $R_{Cu1-Se}$ (Å) | $R_{In1-Se}$ (Å) | $R_{In2-Se}$ (Å) | $R_{Cu2/Vacancy/Li-Se}$ (Å) |
|---|---|---|---|---|
| $CuInSe_2$ | 2.452 | 2.543 | 2.543 | 2.452 |
| $CuIn_2Se_4$ | 2.458 | 2.551 | 2.544 | 2.438 |
| $Cu_{0.5}Li_{0.5}InSe_2$ | 2.492 | 2.587 | 2.574 | 2.445 |

Figure 1

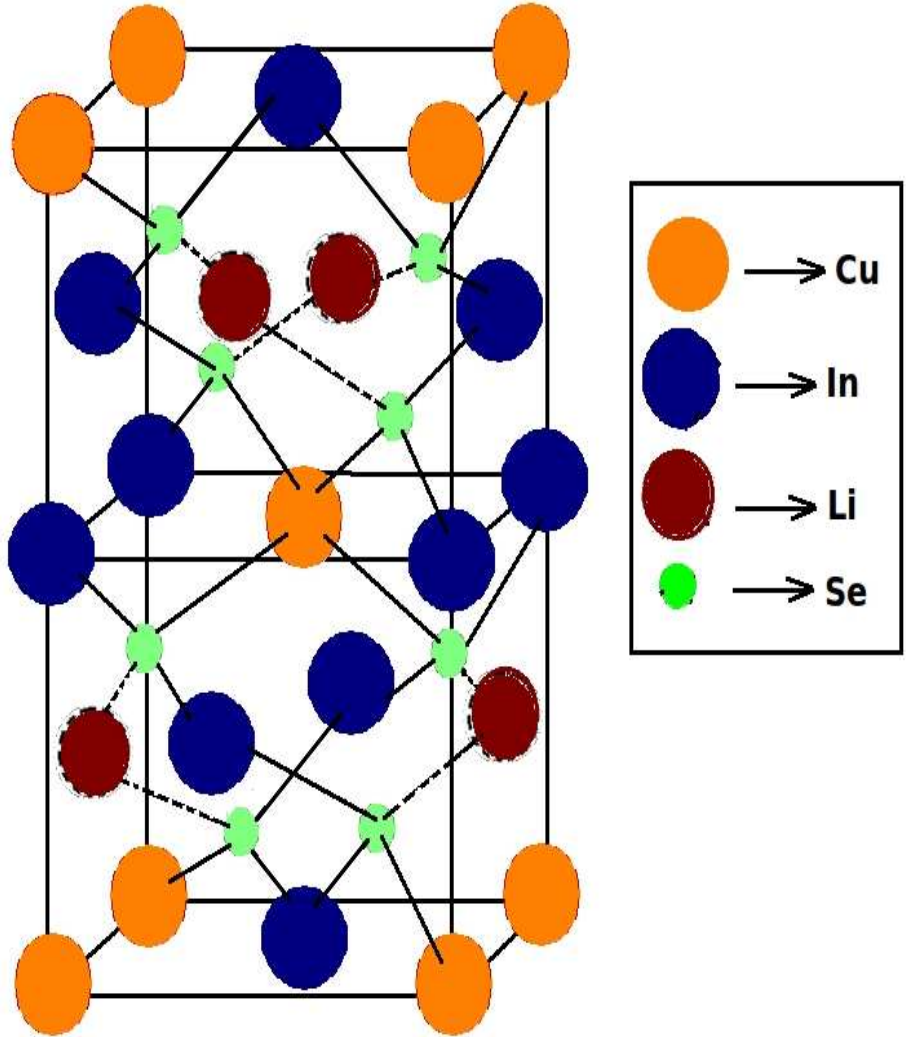

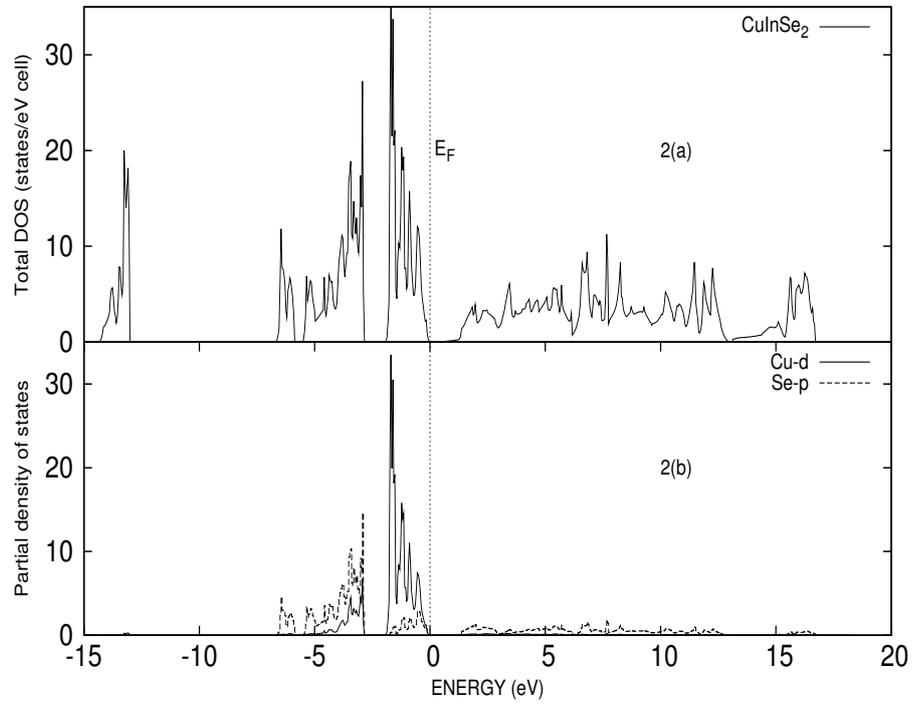

Figure 3

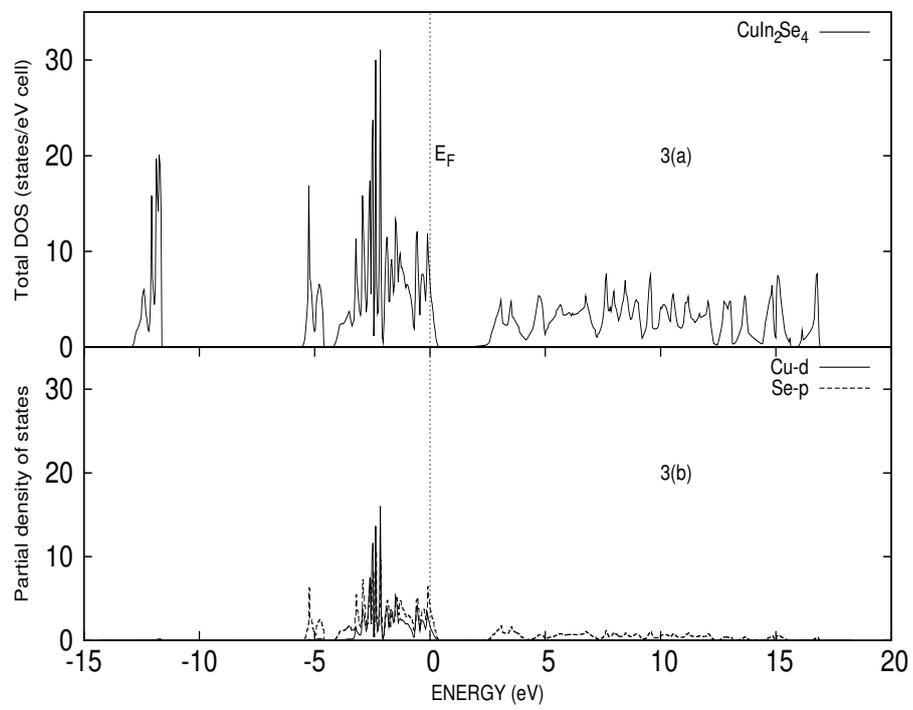

Figure 4

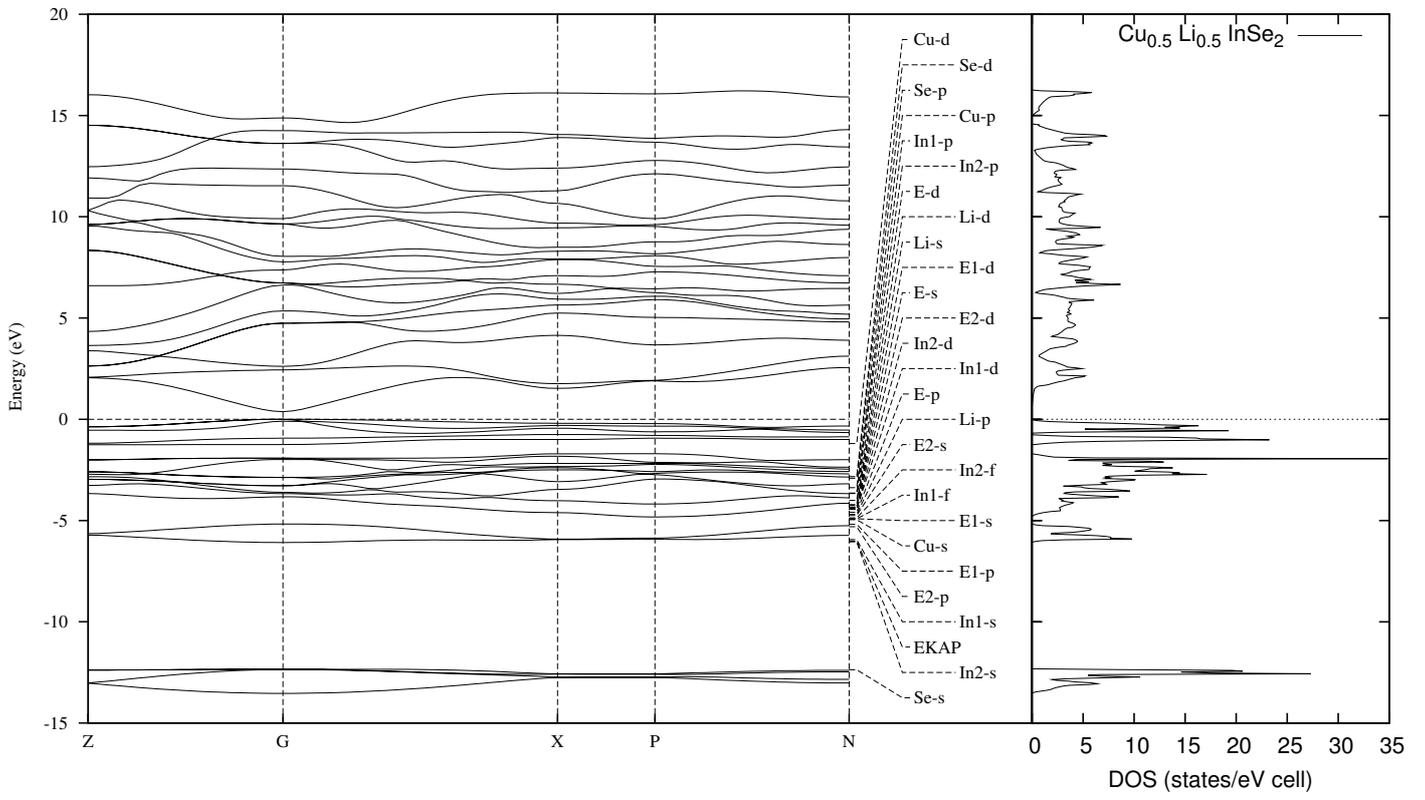

Figure 5

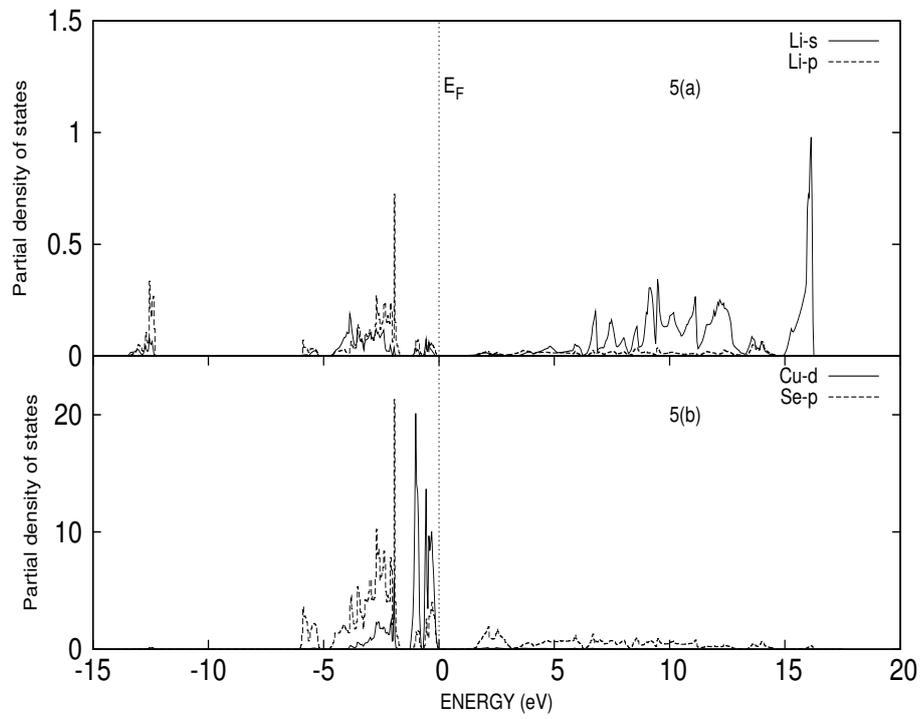

Figure 6

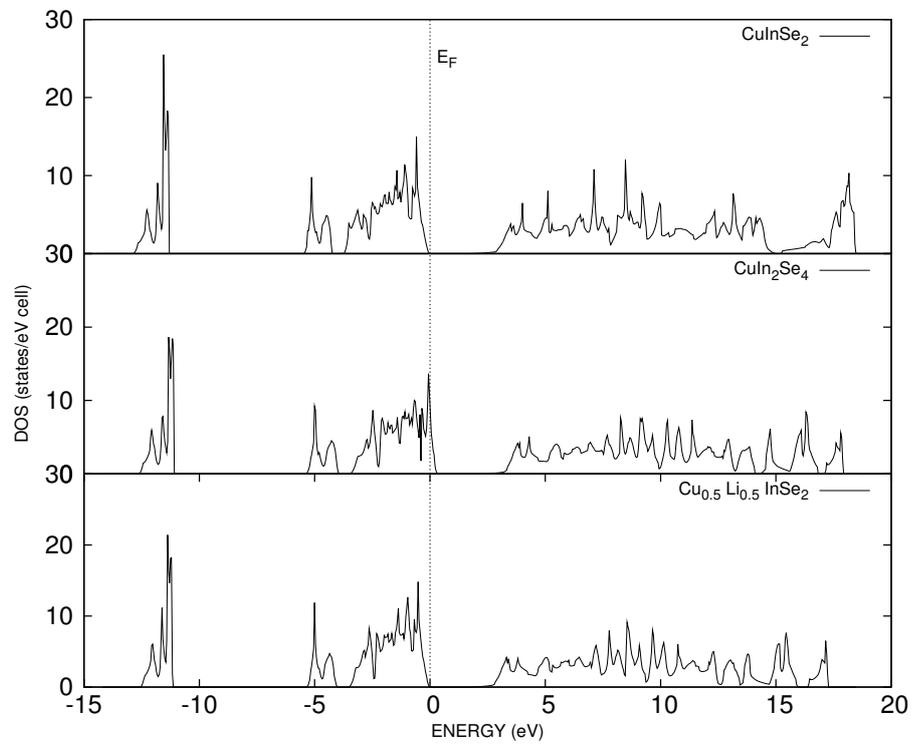

Figure 7

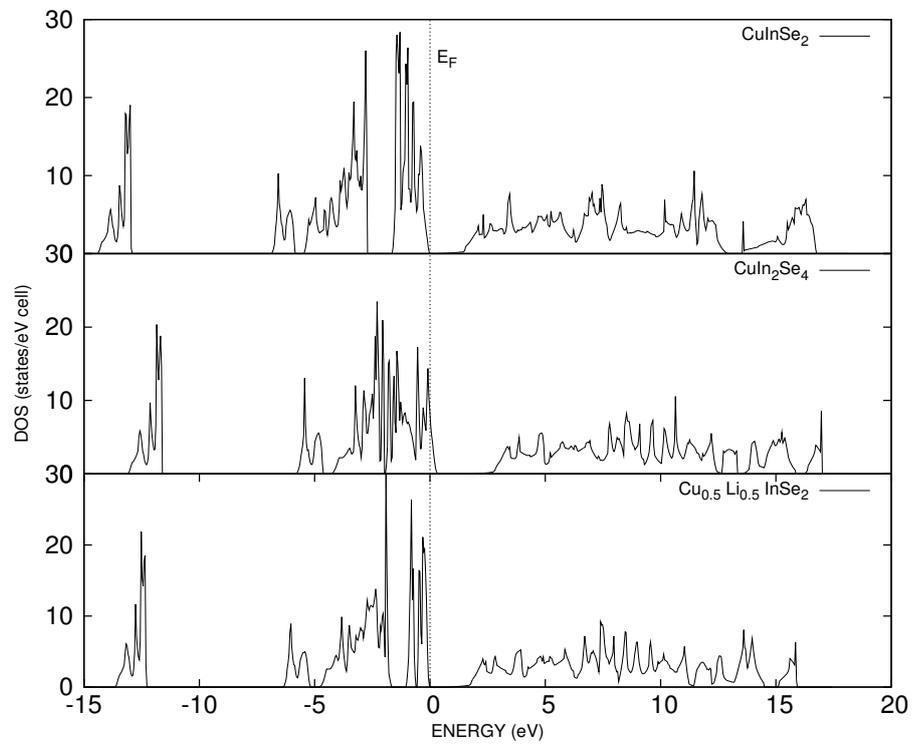